\documentclass[aps,twocolumn,superscriptaddress,showpacs,amsmath,amstex,amssymb,citeautoscript]{revtex4-2}
\usepackage{physics}
\usepackage{siunitx}
\usepackage[colorlinks,linkcolor=magenta,citecolor=magenta]{hyperref}
\usepackage{bm}
\usepackage{graphicx}

\usepackage{xcolor}
\usepackage{lipsum}

\newcommand{\me}{{e}}

\newcommand{\vo}[1]{{\vb{#1}}} 

\newcommand{\up}{\uparrow}
\newcommand{\dn}{\downarrow}

\usepackage{titlesec}

\newcommand{\sectionmain}[1]{\section{#1}.}

\begin{document}
\title{Composite Biased Rotations for Precise Raman Control of Spinor Matterwaves}
\author{Liyang Qiu$^{*}$}
\affiliation{Department of Physics, State Key Laboratory of Surface Physics and Key Laboratory of Micro and Nano Photonic Structures (Ministry of Education), Fudan University, Shanghai 200433, China.}
\email{Current address: Max-Planck-Institut für Quantenoptik, 85748 Garching, Germany.  Email: liyang.qiu@mpq.mpg.de}

\author{Haidong Yuan}
\affiliation{ Department of Mechanical and Automation Engineering, The Chinese University of Hong Kong, Shatin, Hong Kong, China.
}
\author{Saijun Wu}
\affiliation{Department of Physics, State Key Laboratory of Surface Physics and Key Laboratory of Micro and Nano Photonic Structures (Ministry of Education), Fudan University, Shanghai 200433, China.}
 
\begin{abstract}
  Precise control of hyperfine matterwaves via Raman excitations is instrumental to a class of atom-based quantum technology. We investigate the Raman spinor control technique for alkaline atoms in an intermediate regime of single-photon detuning where a choice can be made to balance the Raman excitation power efficiency with the control speed, excited-state adiabatic elimination, and spontaneous emission suppression requirements. Within the regime, rotations of atomic spinors by the Raman coupling are biased by substantial light shifts. Taking advantage of the fixed bias angle, we show that composite biased rotations can be optimized to enable precise ensemble spinor matterwave control within nanoseconds, even for multiple Zeeman pseudo-spins defined on the hyperfine ground states and when the laser illumination is strongly inhomogeneous. Our scheme fills a technical gap in light pulse atom interferometry, for achieving high speed Raman spinor matterwave control with moderate laser power. 
\end{abstract}
    
\maketitle
\section{Introduction}

Unlike microwave control of quantum systems that usually occurs within a sub-wavelength volume with a uniform control intensity~\cite{Harty2014,Rong2015,Wang2016,Zangara2017,Li2017,Starkov2020}, optical controls~\cite{Brif2010,Koch2019} are substantially more prone to intensity inhomogenuities. For  2-level atomic control, the intensity inhomogenuity problem for large samples is further coupled with multi-level perturbations and spontaneous emission, only making the quantum control significantly more difficult to perfect~\cite{Ma2020}. The resulting inaccuracy can bottleneck the advance of quantum technology when high fidelity control of light-atom interaction is required. A particular example is the large-momentum-transfer (LMT)-enhanced atom optical techniques~\cite{McGuirk2000,Muller2009, Kotru2015,Jaffe.2018,Plotkin-Swing2018,Gebbe2021,Rudolph2020,He2020b,Qiu2022,Wilkason2022}, widely expected to drastically enhance the performance and scalability of quantum sensors~\cite{Dimopoulos2008, Morel2020,Wu2020,Greve2021,Overstreet2022},  quantum simulators~\cite{Gross.2017,Daley2022} and  general quantum information processors~\cite{Garcia-Ripoll2003,Duan2004, Mizrahi2013, Heinrich2019}, which requires exceptionally high fidelity to cyclically drive the optical or 2-photon transitions.


Pioneering efforts toward precise control of 2-level atomic ``spinor'' matterwave based on 2-photon Raman excitations were made during the developments of light pulse atom interferometry~\cite{McGuirk2000,Muller2009, Kotru2015,Jaffe.2018}  and ion-based quantum information processing~\cite{Mizrahi2013,Mizrahi2014}. As in the Fig.~\ref{fig:intro}a illustration, here the atomic pseudo-spin is defined on a pair of ground-state hyperfine levels split by $\omega_{\rm HF}$ in frequency, 
referred to as $\ket{\downarrow}$, $\ket{\uparrow}$. The ${\bf k}_{1,2}$ optical pulses with duration $\tau_{\rm c}$, Rabi frequencies $\Omega_{1,2}$ and optical frequency difference $\omega_{1}-\omega_{2}=\omega_{\rm HF}$ resonantly drive the spin-flip while transferring the $\hbar {\bf k}_{\rm eff}$ photon momentum to the spinor matterwave (${\bf k}_{\rm eff}={\bf k}_1-{\bf k}_2$). For short enough $\tau_c$ so that the atomic motion is negligible, the spinor matterwave can be uniformly controlled by a Raman coupling 
$\Omega_{\rm R}$ on the Bloch sphere (Fig.~\ref{fig:intro}c) for realizing {\it e.g.} matterwave splitters and mirrors~\cite{Kasevich.1991}. To ensure a laser intensity-independent 2-photon detuning $\delta$, a moderate single photon detuning $\Delta<\omega_{\rm HF}$ is typically chosen so that the differential Stark shift can be nullified~\cite{Weiss1994}. Separately, for Raman control of microscopically confined ions, THz-level $\Delta$ comparable to the fine-structure splitting $\omega_{\rm fine}$ can be chosen to minimize the Stark shifts~\cite{Wineland2003}. 
Nevertheless, in either case, precise control of large samples is still challenged by the inhomogenuous optical illumination that broadens the $\Omega_{\rm R}$ amplitude, see the Fig.~\ref{fig:optim-pulsearea}a illustration. While efforts have been made to develop adiabatic or composite Raman controls in atom interferometry that are resilient to the intensity errors~\cite{Kotru2015,Berg2015, Jaffe.2018, Saywell2020a,Qiu2022}, practically the benefits are compromised by increased spontaneous emission during the multiple excitations at moderate $\Delta$.

In this Letter, we report systematic development of a method to construct fast composite pulses for Raman control of spinor matterwaves. Our work is motivated by the observation that between the conventional choices of single-photon detuning $\Delta$ for light pulse atom interferometry and trapped-ion quantum information processing, $\omega_{\rm HF}<\Delta\ll \omega_{\rm fine}$, the typically unfavored none-zero Stark shift to the 2-photon detuning $\delta$ is proportional to the Raman Rabi frequency $\Omega_{\rm R}$. The Raman controls  are therefore rotations of atomic spins on the Bloch sphere with the axes biased by a fixed angle $\theta_b$ from the equator (Fig.~\ref{fig:intro}d). The proportionality ensures simple composite strategies~\cite{Wu.2005,Hughes2007,Herold2012} for driving universal qubit gates within $\tau_{\rm c}\sim 100\pi/\omega_{\rm HF}$, typically in nanoseconds so that 2-photon Doppler shifts become negligible~\cite{Berg2015, Saywell2020a,Rudolph2020}.  
Furthermore, since within this biased rotation (BR) regime the errors in the detuning $\delta$ and intensity $I$ are perfectly correlated, simple SU(2) optimization strategy can be applied to achieve precise ensemble control~\cite{Skinner2003,Ruths2012} of large alkaline samples, even with a focused laser beam. Unbounded by traditional choices of single photon detuning $\Delta$~\cite{Weiss1994,Wineland2003}, our scheme conveniently supports adjustable balance of Raman excitation optical power efficiency with the requirements on the control speed and/or the suppression of excited-state dynamics, for realizing {\it e.g.} LMT atom optical control of specific samples~\cite{McGuirk2000, Nelson2020, Gross.2017,Daley2022}, and for achieving precise spinor matterwave control in quantum enhanced atom interferometric applications~\cite{Szigeti2020, Wu2020, Greve2021, Anders2021}.



\section{Biased rotation}

    \begin{figure}[htbp]
        \centering
        \includegraphics[width=\linewidth]{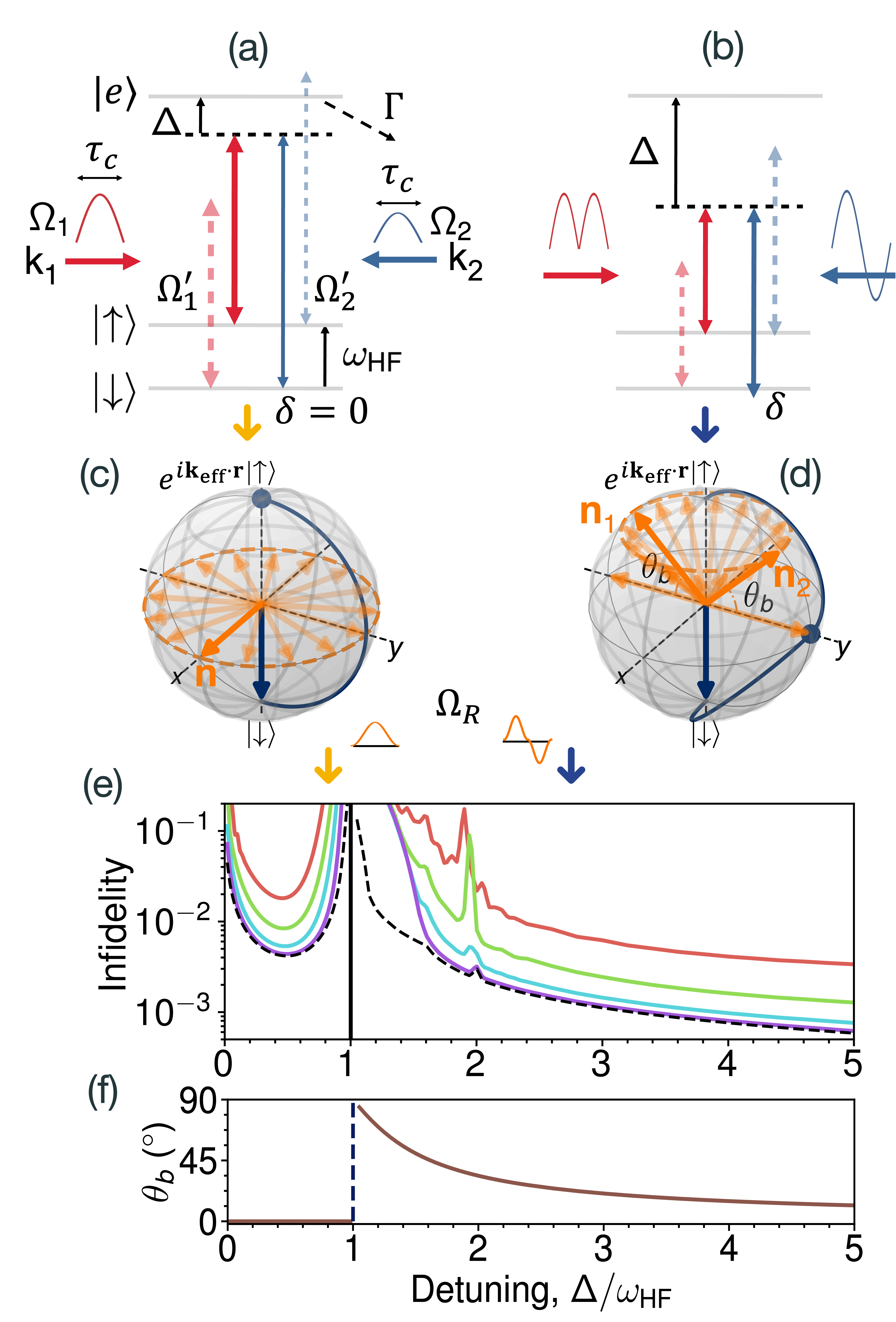}
        \caption{Comparison between the resonant Raman controls in the MR (a,c) and BR (b,d) regime. (a,b): The level diagrams. (c,d): Bloch sphere representation of the optical Raman control under the 2-level approximation. The light orange arrows denote available rotation axes $\vb{n}$ for the generalized Raman Rabi vectors $\va*{\Omega}_{\rm R} = \qty(\Omega_{\rm R}\cos\varphi, \Omega_{\rm R}\sin\varphi, \delta)$. For (c) with $\delta=0$ (MR), the axes are on the equator. For (d) the axes are biased by $\theta_b={\rm tan}^{-1}(\delta/|\Omega_{\rm R}|)$ (BR). Blue lines represent typical spin trajectories during a $\pi$ rotation in the two cases. (e): Gate infidelities $\mathcal{I}$ for the rotation $R_x(\pi)$ vs $\Delta$ with  numerical simulation on the $^{87}$Rb D1 line (Sec.~\ref{sec:num}, with $\omega_{{\rm hfs},e}=0$). Red, green, blue and purple lines represent minimal achievable errors at $\omega_{\rm HF}\tau_{\rm c}/\pi = 20, 50, 100, 200$ respectively. Notice here the $N=2$ double pulse scheme requires $\theta_b<45^{\circ}$, corresponding to $\Delta/\omega_{\rm HF}>1.618$. The features near $\Delta/\omega_{\rm HF}=2$ are due to multi-photon resonances. Fig.~(f) gives the bias angle $\theta_b$ in the Fig.~(e) simulation. 
        }
        \label{fig:intro}
    \end{figure}

We consider the Fig.~\ref{fig:intro}a Raman configuration. The traveling pulses with electric field envelopes $\mathcal{E}_{1,2}e^{i({\bf k}_{1,2}\cdot{\bf r}+\phi_{1,2})}$ and duration $\tau_c$ drive the $\ket{\uparrow}-|e\rangle$ and $\ket{\downarrow}-|e\rangle$ Rabi-couplings $\Omega_{1,2}$ (solid double arrows), as well as the $\Omega_{1,2}'$ (dashed double arrows) with the ground states interchanged. With a large single-photon detuning $\Delta\gg 1/\tau_c$, the excited levels $|e\rangle$ can be adiabatically eliminated
from the ground state dynamics. While both $\Omega_{1,2},\Omega'_{1,2}$ are proportional to $\mathcal{E}_{1,2}$, efficient $\ket{\downarrow}\leftrightarrow \ket{\uparrow}$ Raman coupling can only be induced by $\Omega_{1,2}$ if $\omega_{\rm HF} \gg 1/\tau_c$. The counter-rotating Raman transitions induced by $\Omega_{1,2}'$ are energetically suppressed by the hyperfine splitting. An effective Hamiltonian to describe the ground state spinor can be written as
\begin{equation}
H = -\frac{1}{2}\hbar\delta\sigma_z + \frac{1}{2}\hbar|\Omega_{\rm R}|(\sigma_x\cos\varphi+\sigma_y\sin\varphi).\label{eq:H1}
\end{equation}
Here $\sigma_{x,y,z}$ are Pauli matrices. The Raman Rabi coupling $\Omega_{\rm R}=\Omega_1^*\Omega_2/2\Delta$, with phase $\varphi=\phi_2-\phi_1$, can be designed by shaping the pulse envelopes $\mathcal{E}_{1,2}$.  The $\delta=\delta_{\uparrow}-\delta_{\downarrow}$ in Eq.~(\ref{eq:H1}) is the two-photon detuning by the Stark shifts,
\begin{equation}
    \delta =\frac{|\Omega_1|^2}{4\Delta}+\frac{|\Omega_2'|^2}{4(\Delta-\omega_{\rm HF})}-\frac{|\Omega_2|^2}{4\Delta}-\frac{|\Omega_1'|^2}{4(\Delta+\omega_{\rm HF})}.\label{eq:shift}
\end{equation}

\begin{figure*}[htbp]
        \centering
        \includegraphics[width=1 \linewidth]{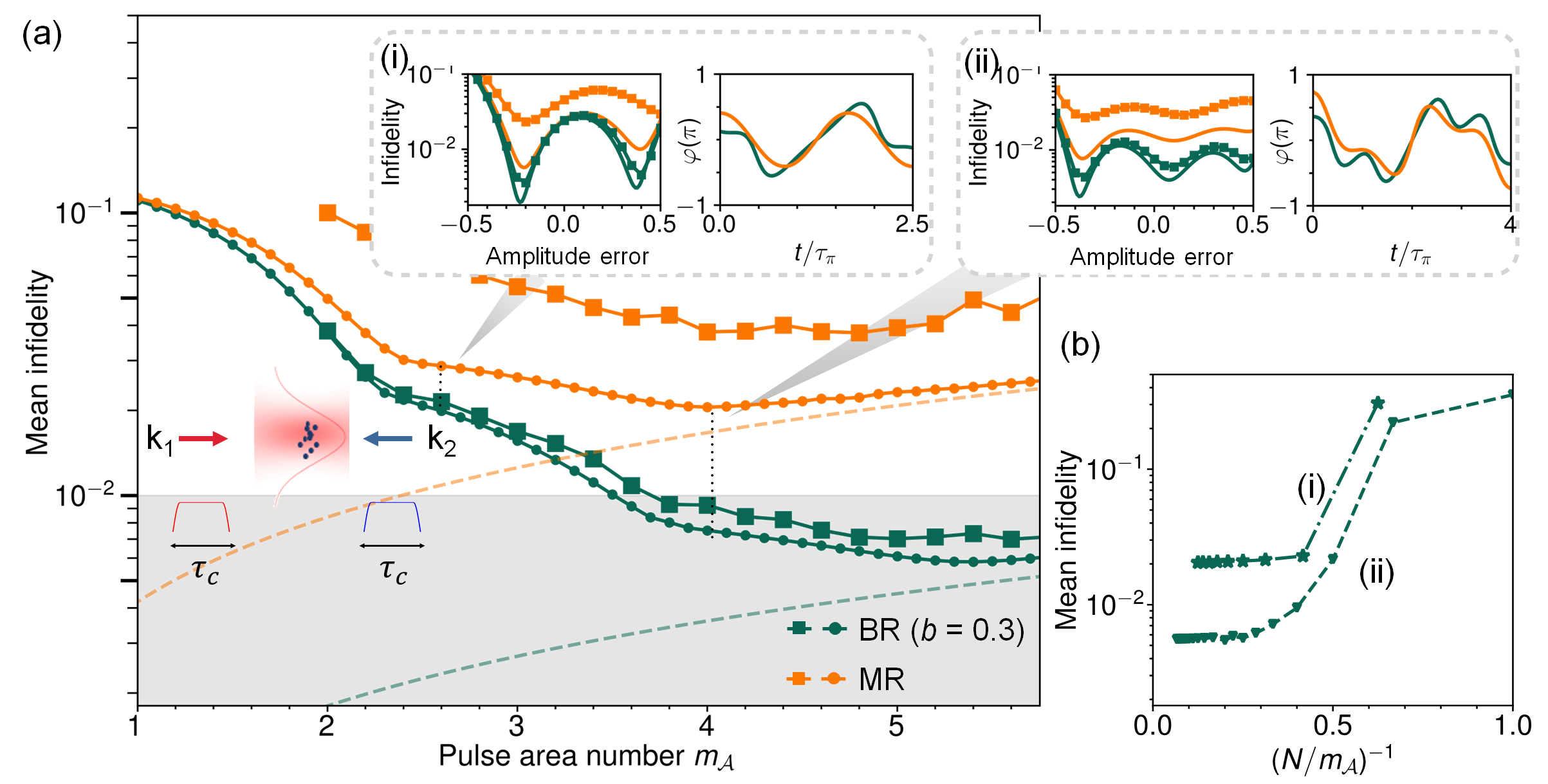}
        \caption{(a) Performance of a $R_x(\pi)$ gate with CP-BR (b=0.3, $\Delta=3.6\omega_{\rm HF}$, blue) and CP-MR (b=0.0, $\Delta=0.5\omega_{\rm HF}$, orange). The infidelity $\overline{\mathcal{I}}$ averaged over $\varepsilon_{\mathcal{A}}/\mathcal{A}=\pm 50\%$ is plotted vs $m_{\mathcal{A}}$. The dashed lines give the Eq.~(\ref{eq:Isp}) spontaneous emission limit. The top insets give typical $\mathcal{I}$ vs $\varepsilon_{\mathcal{A}}/\mathcal{A}$, before the ensemble average, by the optimal pulse with phase profile $\{\varphi_j\}_{\rm opt}$ (before the $|\Omega_{\rm R}(t)|$ edge-smoothing), with $m_\mathcal{A}=2.5$ (i) and $m_\mathcal{A}=4$ (ii) respectively. The square symbols give $\overline{\mathcal{I}}$ according to full-level $^{87}{\rm Rb}$ D-line simulation, with the SU(2)-optimized $\{\varphi_j\}_{\rm opt}$ and edge-smoothed $|\Omega_{\rm R}(t)|$, $\tau_c=(m_{\mathcal{A}}+\pi/2-1)\tau_{\pi}$, with $\tau_{\pi}=100/\omega_{\rm HF}=7.4$~ns. (b) $\overline{\mathcal{I}}$ vs sub-pulse number $N$, evaluated with 2-level model, with other parameters by  (a,i) and (a,ii).}
        \label{fig:optim-pulsearea}
    \end{figure*}

For short enough $\tau_c$~\cite{foot:frozen} and with spatially uniform intensities $I_{1,2}=|\mathcal{E}_{1,2}|^2$~\cite{foot:phase}, the delocalized spinor matterwave $|\psi({\bf r})\rangle= \psi_{\downarrow}({\bf r})\ket{\downarrow}+\psi_{\uparrow}({\bf r}) e^{i {\bf k}_{\rm eff}\cdot {\bf r}} \ket{\uparrow}$ can be uniformly controlled with a single set of $(\delta,\Omega_{\rm R})$ parameters by Eq.~(\ref{eq:H1}). Practically, for achieving ultra-precise matterwave control with a precision similar to those for internal spins~\cite{Harty2014,Ballance2016, Wang2016}, particularly for large samples with inhomogenuous $I_{1,2}$ (See Fig.~\ref{fig:optim-pulsearea}a inset), then the spinor matterwave control must be achieved in an intensity-insensitive manner. 
In particular, the dependence of $\delta$ on $I_{1,2}$ can be suppressed at moderate $\Delta < \omega_{\rm HF}$, which we refer to as a ``magic ratio'' (MR) regime (Fig.~\ref{fig:intro}(a,c)), by tuning the $I_1/I_2$ ratio~\cite{Weiss1994}.

In this work, instead of minimizing the dependence of $\delta$ on $I_{1,2}$~\cite{Weiss1994,Wineland2003,Mizrahi2014}, we focus on Raman spinor matterwave control within $\omega_{\rm HF}<\Delta\ll \omega_{\rm fine}$ so the 2-photon shift $\delta$ is substantial. Correspondingly, the spin rotation axes as in Fig.~\ref{fig:intro}d are biased from the equator of the Bloch sphere by 
\begin{equation}
    \theta_b={\rm tan}^{-1}(\delta/|\Omega_{\rm R}|).\label{eq:thetab}
\end{equation}
Although the $\theta_b$-bias appears inconvenient, for matterwave Raman control with uniform $I_{1,2}$, the bias is easily compensated for by simple composite pulses~\cite{Wu.2005,Hughes2007,Herold2012}. The simplest idea of double-pulse~\cite{Wu.2005} BR is illustrated in Fig.~\ref{fig:intro}d. Here, as long as $\theta_b<\pi/4$, then a double-pulse Raman control can be synthesized for $R_x(\pi)$, a $\pi$-rotation of spin along the $x-$axis. The first sub-pulse with Raman ``Rabi vector'' $\vec{\Omega}_{\rm R}=(|\Omega_{\rm R}|\cos\varphi,|\Omega_{\rm R}|\sin\varphi,\delta)$ transports the state vector to the equator during $0<t<\tau_c/2$. The 2nd sub-pulse, with $\varphi\rightarrow \varphi+\pi$ realized by introducing a relative $\pi$ phase-jump between $\mathcal{E}_{1,2}$ (Fig.~\ref{fig:intro}b), completes the spin inversion during $\tau_c/2<t<\tau_c$. 


To illustrate the BR advantages in real atoms, in Fig.~\ref{fig:intro}e we compare the performance of the exemplary $R_x(\pi)$ rotation ($U=-i\sigma_x$) on the $^{87}$Rb D1 line (Fig.~\ref{fig:level})~\cite{Steck2010} in the MR regime ($\Delta<\omega_{\rm HF}$, with suitable $I_1/I_2$  to nullify $\delta$~\cite{Weiss1994}) with those achievable in the BR regime ($\Delta>\omega_{\rm HF}$ with a fixed $I_1/I_2=1$). The gate fidelity is defined as
\begin{equation}
    \mathcal{F}=\frac{1}{6}\sum_{j=1}^6|\langle \psi_j| U^{\dagger}\tilde U|\psi_j\rangle|^2.\label{eq:gatef}
\end{equation}
Here $U$ is the target operator in the $\{\ket{\uparrow},\ket{\downarrow}\}=\{\ket{\uparrow_0}, \ket{\downarrow_0}\}$ spin-space, defined on the $^{87}$Rb 5S$_{1/2}$ clock transition (Fig.~\ref{fig:level}) for now. The $\tilde U=\hat P_{\sigma}\hat T e^{-i/\hbar \int_0^{\tau_c} H_{\rm eff}{\rm d}t}\hat P_{\sigma}$ is the actual D1 evolution operator projected to the spin space ( $\hat P_{\sigma}$ the projection operator), evaluated by integrating the Schr\"odinger equation with the non-Hermitian Hamiltonian $H_{\rm eff}=H_0-i\frac{\hat \Gamma}{2}$. The $H_0$ describes the full vectorial laser-atom interaction, while $\hat \Gamma$ describes the spontaneous emission from $|e\rangle$, detailed in Sec.~\ref{sec:num}.  The gate fidelity is obtained by averaging a set of $\{|\psi_j\rangle, j=1,..,6\}$ initial states, which are the three pairs of eigenstates for $\sigma_{x,y,z}$~\cite{Magesan2012}. 

In the Fig.~\ref{fig:intro}e example the optical pulses with duration $\tau_c$ are sine-shaped to minimize non-adiabatic $|e\rangle$ excitations. To simplify the analysis, the excited state hyperfine splitting $\omega_{{\rm hfs},e}$ is set as zero to fully suppress the spin-leakages among the Zeeman sub-levels for the linearly polarized pulses~\cite{Qiu2022} (Sec.~\ref{sec:num}). The infidelity  $\mathcal{I}=1-\mathcal{F}$ degrades at short $\tau_{\rm c}$, since the shorter $\tau_c$ is associated with a stronger spectra component to excite  $|e\rangle$. By increasing $\tau_c$ to $100\pi/\omega_{\rm HF}$ and for, {\it e.g.}, $\Delta>2.5 \omega_{\rm HF}$, non-adiabatic excitations to $|e\rangle$ including those associated with multi-photon resonances are effectively suppressed. In this $|e\rangle-$elimination limit, the gate dynamics follows the Eq.~(\ref{eq:H1}) SU(2) model (Fig.~\ref{fig:intro}(c,d)) with the infidelity $\mathcal{I}$ reaching a minimal spontaneous emission error $\mathcal{I}_{\rm sp}$ associated with the dressed ground states,
\begin{equation}
\mathcal{I}_{\rm sp} =\eta \mathcal{A} \Gamma/\Delta.\label{eq:Isp}
\end{equation}
Here $\mathcal{A}=\int_0^{\tau_c}|\Omega_{\rm R}|{\rm d}\tau$ is the Raman pulse area for completing $R_x(\pi)$. The factor $\eta$ of order unity depends on the detuning $\Delta$ and the associated choice of $I_1/I_2$ ratio (dashed lines in Fig.~\ref{fig:intro}e, evaluated by increasing $\tau_c$ to $1000\pi/\omega_{\rm HF}$).  For this fictitious $^{87}$Rb example, the Fig.~\ref{fig:intro}e results clearly demonstrate the BR advantages, with $\mathcal{F}>99.8\%$ achievable at $\tau_c=100\pi/\omega_{\rm HF}$ and $\Delta>2.5\omega_{\rm HF}$,  which cannot be reached in the traditional MR regime even in the long-$\tau_c$ limit. We note this BR-to-MR advantage is more pronounced with increased $\Gamma/\omega_{\rm HF}$ ratio in lighter alkalines, and when $\omega_{{\rm hfs},e}$ is included as to be discussed with Fig.~\ref{fig:optim-pulsearea}.

\section{Composite-BR with designed robustness}\label{sec:errrescontrol}


Practically, as illustrated in the inset of Fig.~\ref{fig:optim-pulsearea}a, the optical Raman control of large samples is prone to $\Omega_{\rm R}\propto \sqrt{I_1 I_2}$ amplitude errors associated with intensity inhomogenuity. For the pulsed controls, the amplitude error is characterized by a distribution of deviation $\{\varepsilon_{\mathcal{A}}\}$ from the designed Raman pulse area $\mathcal{A}$. Separately, the 2-photon detuning $\delta$ in Eq.~(\ref{eq:H1}) can also be broadened~\cite{Weiss1994}, such as by the 2-photon Doppler shift if the atomic sample has substantial velocity distribution along ${\bf k}_{\rm eff}$~\cite{Dunning.2014jm,Saywell2020a, Wilkason2022}. The common resolution to suppressing the errors associated with the $(\varepsilon_{\mathcal{A}}, \varepsilon_{\delta})$ deviations relies on ensemble optimal control techniques~\cite{Skinner2003,Ruths2012}, which can be implemented with composite pulses (CP)~\cite{Levitt1986,Li2006a,Dunning.2014jm,Saywell2020a}.

\begin{figure}[htbp]
    \centering
    \includegraphics[width=1\linewidth]{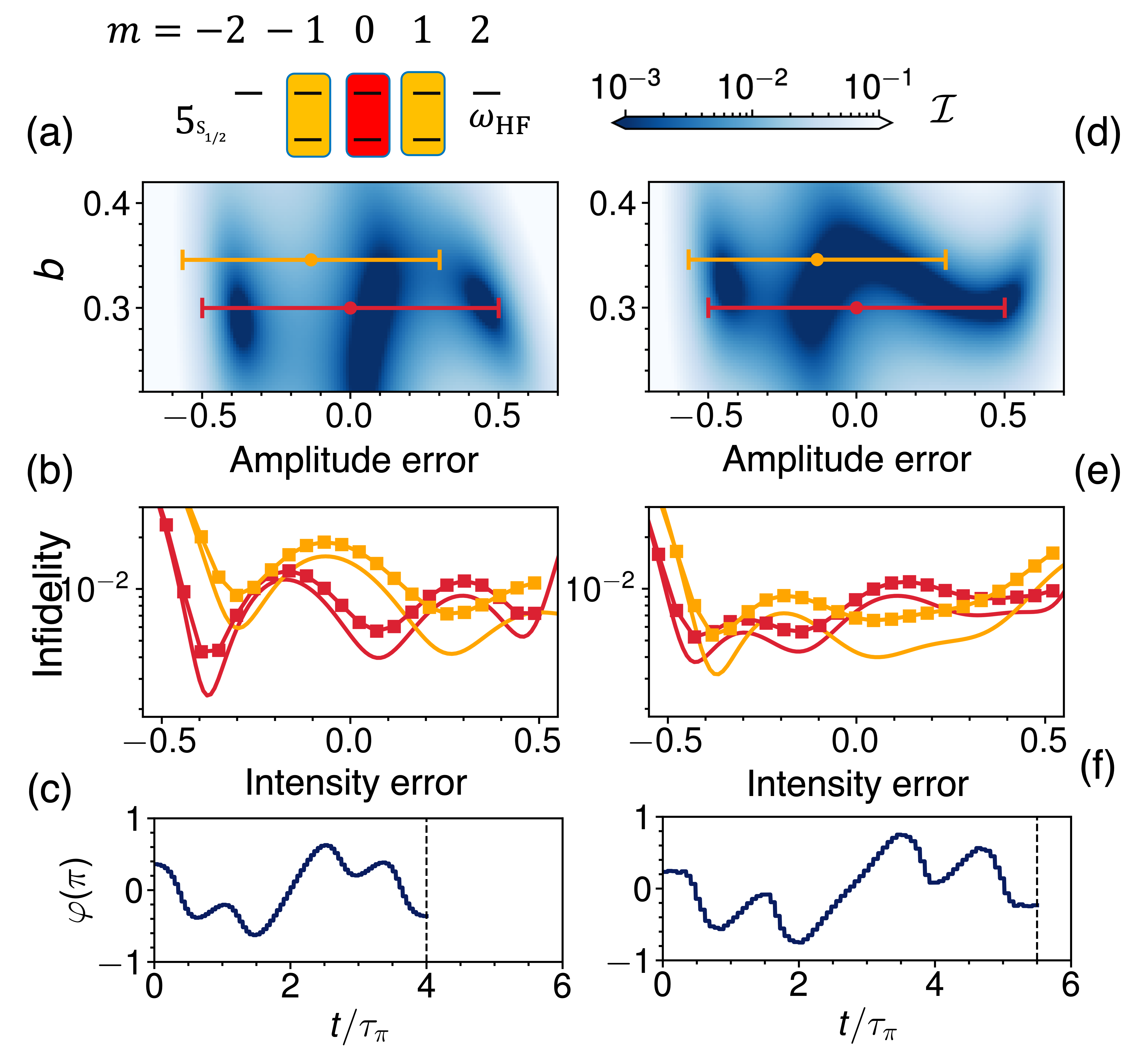}
    \caption{CP-BR for $R_x(\pi)$ to address multiple hyperfine atomic spins. (a,d): Performance of ensemble optimized CP-BR in the parameter space: $\mathcal{I}(\mathcal{A}+\varepsilon_{\mathcal{A}},b)$ vs $(\varepsilon_{\mathcal{A}}/\mathcal{A},b)$, with $\mathcal{A}=4\pi, 5.5\pi$ for Fig.~(a)(d) respectively. 
    The red and yellow bars suggest the range of $(\varepsilon_{\mathcal{A}}/\mathcal{A},b)$ for the ensemble control of $^{87}\mathrm{Rb}$ $m=0$ and  $m=\pm 1$ sub-spins respectively. The level diagram on the top of Fig.~(a) marks the $\{\ket{\uparrow_m},\ket{\downarrow_m}\}$ sub-spins defined on the $^{87}$Rb ground states. The optimization in Figs.~(a-c) only averages $\varepsilon_{\mathcal{A}}$ along the red bar ($m=0$ with $b=0.3$). Instead, the optimization in (d-f) averages $\{\varepsilon_{\mathcal{A}},b\}$ combinations along both the red ($m=0$ with $b=0.3$) and yellow ($m=\pm 1$ with $b=0.35$) bars. (b,e): $\mathcal{I}$ vs $\varepsilon_{I}/I$ for $m=0$ (red) and $m=\pm1$ (yellow) sub-spins. The solid lines and square symbols are from the 2-level and full-model simulations respectively. (c,f): The optimal phase profiles $\qty{\varphi_j}_{\rm opt}$ ($N=80$, before the $|\Omega_{\rm R}(t)|$ edge-sine-smoothing)}
    \label{fig:optim2d}
\end{figure}

Here, with $\tau_c\sim 100\pi/\omega_{\rm HF}$ within tens of nanoseconds for typical alkalines, the errors induced by the 2-photon Doppler shifts are negligible in cold atomic samples~\cite{Kotru2015,Jaffe.2018,Qiu2022} and collimated atomic beams~\cite{McGuirk2000}. The Stark-shifted $\delta =b|\Omega_{\rm R}|$ is proportional to $|\Omega_{\rm R}|$ with the bias ratio $b={\rm tan}\theta_b$ . At a fixed $I_1/I_2$ and for a fixed pair of atomic levels to form the $\{\ket{\uparrow},\ket{\downarrow}\}$ pseudo-spin (Fig.~\ref{fig:level}, Fig.~\ref{fig:optim2d}a), then the $\varepsilon_{\delta}$ and $\varepsilon_{\mathcal{A}}$ errors are perfectly correlated to support a simpler CP-BR strategy. To optimize the robustness of CP-BR, we split an $\mathcal{A}$-area Raman pulse into $N$ 
 equiangular sub-pulses~\cite{Low2016} with Raman phases $\varphi_j$. The full evolution operator 
\begin{equation}\label{eq:Ucp}
    \tilde U^{(N)}(\mathcal{A},b;\{\varphi_j\}) = \prod_j^N \tilde U(\mathcal{A}/N,b;\varphi_j)
\end{equation}
becomes a product (multiply from left) of the single-pulse propagators $\tilde U_j=\tilde U(\mathcal{A}/N,b;\varphi_j)$. We then optimize the fidelity  averaged over the list of $\{\varepsilon_{\mathcal{A}}\}$ deviation of interest for realizing certain operation $U$
\begin{equation}
    \overline{\mathcal{F}}(\mathcal{A},b;\{ \varphi_j \})=\left\langle \mathcal{F}^{(N)}(\mathcal{A}+\varepsilon_{\mathcal{A}}, b;\{ \varphi_j \})\right\rangle_{\{ \varepsilon_{\mathcal{A}}\}}.    \label{eq:cost}
\end{equation}
Here $\mathcal{F}^{(N)}(\mathcal{A}+\varepsilon_{\mathcal{A}}, b;\{\varphi_j\})$ is evaluated according to Eq.~(\ref{eq:gatef}), but with $\tilde U$ replaced by $\tilde U^{(N)}(\mathcal{A}+\varepsilon_{\mathcal{A}},b;\{\varphi_j\})$. In this work, benefited from the simple $\tilde U (\mathcal{A}/N,b;\varphi_j)$ expression (Eq.~\eqref{eq:propagtor-single}), the GRAPE (GRadient Ascent Pulse Engineering) optimization (Sec.~\ref{sec:numoptim})~\cite{Khaneja2005} is performed at the SU(2) level with Eq.~(\ref{eq:H1}) first. The optimal $\{ \varphi_j \}_{\rm opt}$ are then transferred to the full model (Eq.~(\ref{eq:HD1})) to validate the applicability of the CP-BR to real atoms. 




An example of optimal $\overline{\mathcal{I}}_{\rm opt}(\mathcal{A})=1-\overline{\mathcal{F}}_{\rm opt}(\mathcal{A})$ for $R_x(\pi)$ with improved amplitude-error resilience is shown in Fig.~\ref{fig:optim-pulsearea}a vs pulse area number $m_{\mathcal{A}}=\mathcal{A}/\pi$ (disk symbols). Here, for the Eq.~(\ref{eq:cost}) optimization, the amplitude deviation uniformly samples $\varepsilon_{\mathcal{A}}\in(-0.5,0.5) \mathcal{A}$.  The bias ratio is set as  $b=0.0$ (MR, with $\Delta=0.5\omega_{\rm HF}$) and $b=0.3$ (BR, with  $\Delta=3.6\omega_{\rm HF}$). As in the Fig.~\ref{fig:optim-pulsearea}b examples, we typically find $m_{\mathcal{A}}/N\leq 0.3$ to be enough for CP to reach the optimal performance. On the other hand, to help mitigating non-adiabatic $|e\rangle$ excitation, we set large enough $N=80$ sub-pulses so that the optimal $\{\varphi_j\}_{\rm opt}$ can become quasi-continuous in time. The phase symmetry in Fig.~\ref{fig:optim-pulsearea}(a,i-ii) is associated with the $x-$rotation~\cite{Kobzar.2012}. With Eq.~(\ref{eq:Isp}), we include the minimal $\mathcal{I}_{\rm sp}$ into the SU(2) $\overline{\mathcal{I}}_{\rm opt}$, after straightforward light intensity average.

From Fig.~\ref{fig:optim-pulsearea}a we observe increasingly efficient suppression of the average infidelity $\overline{\mathcal{I}}_{\rm opt}(\mathcal{A})$ with larger pulse area number $m_{\mathcal{A}}$, which is expected since redundant rotations on the Bloch sphere can be phased to improve the control robustness~\cite{Levitt1986,Zhu2002, Li2006a, Ichikawa.2012,Barnes2015}. 
Interestingly, the enhancement to $\overline{\mathcal{I}}_{\rm opt}(\mathcal{A})$ with $\mathcal{A}$ displays step-wise features near certain integers $m_{\mathcal{A}}$, which are associated with increased number of perfected $\mathcal{I}(\mathcal{A})$ within the $\pm 50\%$ $\varepsilon_{\mathcal{A}}/\mathcal{A}$ distribution, as shown by the Fig.~\ref{fig:optim-pulsearea}a(i,ii) examples. The features could merit additional study in the future. Here, we note that increasing the average $\mathcal{A}$ leads to increased $I_{\rm sp}$ to limit the achievable $\overline{\mathcal{I}}_{\rm opt}$. For the particular ensemble control with the required error resilience, a balance is met at a suitable $\mathcal{A}_0$ for $\overline{\mathcal{I}}_{\rm opt}(\mathcal{A}_0)$ to reach its minimum. For the $\overline{\mathcal{I}}_{\rm opt}$ at the SU(2) level, we already see the $R_x(\pi)$ gate in the BR regime performs substantially better when addressing the $\varepsilon_{\mathcal{A}}/\mathcal{A}=\pm 50\%$ amplitude distribution, with BR-$\overline{\mathcal{I}}_{\rm opt}(\mathcal{A}_0)\approx 0.5\%$ reached at $\mathcal{A}_0\approx 5.5\pi$, as compared to MR-$\overline{\mathcal{I}}_{\rm opt}(\mathcal{A}_0)\approx 2\%$ at $\mathcal{A}_0\approx 4\pi$, due to the better suppression of spontaneous emission.

We now apply the SU(2)-optimized $\{\varphi_j\}_{\rm opt}$ to the full $^{87}$Rb D1 model. As detailed in Sec.~\ref{sec:num}, to mitigate non-adiabatic excitations to the $|e\rangle$ states during this step, $|\Omega_{\rm R}|\propto \sqrt{I_1I_2}$ is edge-sine-smoothed from a $\tau_c=m_{\mathcal{A}}\tau_{\pi}$ square pulse into a $\tau_c=(m_{\mathcal{A}}+\pi/2-1)\tau_{\pi}$ pulse, before the splitting into the $N$ equal-area sub-pulses to associate with the quasi-continuous $\{\varphi_j\}_{\rm opt}$. Since $\delta\propto|\Omega_{\rm R}|$, the reshaped Raman pulse retains the full control advantages at the SU(2)-level by Eq.~(\ref{eq:H1}).  In light of the Fig.~\ref{fig:intro}e knowledge, we set the ``$\pi$ time'' $\tau_{\pi}=100\pi/\omega_{\rm HF}$ for the Fig.~\ref{fig:optim-pulsearea} simulation, with square symbols to represent the associated BR-$\overline{\mathcal{I}}_{\rm opt}$ and MR-$\overline{\mathcal{I}}_{\rm opt}$.


In Fig.~\ref{fig:optim-pulsearea} we find the BR-$\overline{\mathcal{I}}_{\rm opt}$ in the full model only degrades slightly from that by the 2-level model. 
In contrast, the full model MR-$\overline{\mathcal{I}}_{\rm opt}$ (yellow square symbols) degrades substantially from the SU(2) prediction. Similar to Fig.~\ref{fig:intro}e, part of the degraded performance in both cases is associated with non-adiabatic $|e\rangle$ excitations. In addition, with the $\omega_{{\rm hfs},e}=0.12\omega_{\rm HF}$ excited state hyperfine
splitting for $^{87}$Rb~\cite{Steck2010} restored, the $\Delta m=\pm 2$ spin leakage~\cite{Happer1972} is retained at a $\omega_{{\rm hfs},e}/\Delta$ level, during the $\Delta m=0$ Raman excitation driven by linearly polarized pulses (Eq.~(\ref{eq:leakrabi}))~\cite{Qiu2022}. Due to the moderate $\Delta=0.5\omega_{\rm HF}$, the spin leakages more severely affect MR-$\overline{\mathcal{I}}_{\rm opt}$.  We note that within nanosecond $\tau_c$, the spin-leakage among the Zeeman sublevels cannot be easily suppressed by the traditional bias-field method that sufficiently lifts the Zeeman degeneracy~\cite{Weiss1994,    Kasevich1992}. A CP-BR-compatible, large enough $\Delta\gg\omega_{{\rm hfs},e}$ is therefore important for isolating the target atomic spins (Fig.~\ref{fig:level}, Fig.~\ref{fig:optim2d}a) during the fast spinor matterwave control.



We finally consider the bias angle $\theta_{b}$ (Eq.~(\ref{eq:thetab})) variations. As illustrated in Fig.~\ref{fig:level} with the $^{87}$Rb example, for a generic alkaline atom with $I>1/2$ nuclear spin and when addressing $\{\ket{\uparrow_m},\ket{\downarrow_m}\}$ pseudo-spins defined on the $F=I\pm 1/2$ hyperfine levels (Fig.~\ref{fig:optim2d}a), with $m=m_F$, then both the Raman Rabi frequency $\Omega_{{\rm R},m}$ and the bias ratio $b_m$ becomes $m$-dependent~\cite{Dunning.2014jm,Qiu2022}.  We consider the linearly polarized composite pulse design to address all the $\{\ket{\uparrow_m},\ket{\downarrow_m}\}$ sub-spins of $^{87}$Rb. 
The optimization to achieve resilience against laser intensity error $\varepsilon_I/I\in (-0.5, 0.5)$ (Fig.~\ref{fig:optim2d}(b,e))
is again performed at the SU(2) level first to obtain $\{\varphi_j\}_{\rm opt}$, which is then transferred to the full model. Figure~\ref{fig:optim2d}(a-c) results are optimized for the $m=0$ sub-spin only, covering $\varepsilon_{\mathcal{A}}\in (-0.5,0.5)\mathcal{A}$ with $\mathcal{A}=4\pi$ at $b=0.3$. In contrast, Figure~\ref{fig:optim2d}(d-f) results are optimized to balance the performance for all the $m=0,\pm 1$ sub-spins, covering $\varepsilon_{\mathcal{A}}\in (-0.5,0.5)\mathcal{A}$ with $\mathcal{A}=5.5\pi$ at $b=0.3$ ($m=0$) and $\mathcal{A}=4.7\pi$ at $b=0.35$ ($m=\pm 1$). The parameter coverages are marked with red ($m=0$) and yellow ($|m|=1$) lines in the Fig.~\ref{fig:optim2d}(a,d) parameter-space 2D plot of $\mathcal{I}$. The optimized phase $\{\varphi_j\}_{\rm opt}$ for the composite pulses are given in Fig.~\ref{fig:optim2d}(c,f). 
The full model results in Fig.~\ref{fig:optim2d}(b,e) are again with the pulse profile edge smoothed at $\tau_{\pi}=100\pi/\omega_{\rm HF}$), as those for Fig.~\ref{fig:optim-pulsearea}. 
According to Fig.~\ref{fig:optim2d}e, we find $\overline{\mathcal{I}}_{\rm opt}\approx 9\times 10^{-3}$ for the composite $R_x(\pi)$ gate when addressing all the $^{87}$Rb sub-spins with the $\pm 50 \%$ laser intensity distribution. This performance is only slightly compromised from  the single $m=0$ sub-spin result in Fig.~\ref{fig:optim-pulsearea}a where $\overline{\mathcal{I}}_{\rm opt}(5.5\pi)\approx 7\times 10^{-3}$ is reached. The overall performance is improvable further with larger $\mathcal{A}$ by increasing $\Delta$ in proportion to maintain a low spontaneous emission level.

\section{Discussion and Outlook}\label{sec:conclusion}

From inertial and gravity sensing~\cite{Peters1997, Gustavson2000, Hu2013,Overstreet2022} to quantum simulation~\cite{Gross.2017,Daley2022} and computation~\cite{Garcia-Ripoll2003,Duan2004, Mizrahi2013,Lo2015,Fluhmann2019, Heinrich2019}, atom interferometric sensing of spin-and-spatial dependent interaction is highly useful for atom-based quantum technology. Operating the interferometry technique in the quantum regime~\cite{Szigeti2020, Wu2020, Greve2021, Anders2021} requires high-fidelity coherent control of spinor matterwave. However, for generic reasons associated with phase-space density limitation, atomic wavefunction and ballistic expansion, as well as the imperfect optical field collimation itself, free-space matterwave can hardly be sufficiently localized in space to be immune to inhomogenuous optical field broadening. To this end, the composite pulse technique originally developed in NMR~\cite{Levitt1986,Skinner2003,Ruths2012,Li2006a,Low2016} becomes particularly useful for realizing
error-resilient ``spinor matterwave gates'' with light, for achieving ultra-high control fidelity. However, a prerequisite for achieving the goal is to rapidly manipulate the isolated pseudo-spins defined on pairs of atomic levels, before any decoherence occur, and even before the atoms move.  

This work revisits spinor matterwave control techniques based on Raman excitations. Our work focuses in a regime of Raman control with $\Delta=\mathcal{O}(\omega_{\rm HF})$ that supports high-speed operation with $\tau_c\sim 100\pi/\omega_{\rm HF}$ within nanoseconds, only limited by the ground state hyperfine splitting. This choice of $\Delta$ leads to substantial light shift to the 2-photon detuning $\delta$. However, we have shown that  taking advantage of perfect correlation between the $\delta$ and $\Omega_{\rm R}$ errors, composite biased rotations can be optimized for precise ensemble spinor matterwave control, even for multiple Zeeman pseudo-spins and when subjected to inhomogeneous laser illumination. Related to the underlying geometric robustness~\cite{Kabytayev2014,Barnes2015}, we find the optimal CP-BR to be fairly tolerant to the pulse parameter errors themselves too (Sec.~\ref{sec:robust}).  We also notice that beyond error suppression, CP-BR can be designed to enhance the parameter selectivity~\cite{Low2016}, such as for improving the spatial resolution when addressing arrays of samples~\cite{Bluvstein2022}. 

So far, we have ignored the common Stark shift $\delta_{\rm com}=(\delta_{\uparrow}+\delta_{\downarrow})/2$ to the spinor matterwave dynamics~\cite{Weiss1994}. In absence of spin-motion separation, {\it i.e.}, with $|\psi_{\uparrow}({\bf r})|^2=|\psi_{\downarrow}({\bf r})|^2$, then the common Stark shift does not affect the spinor coherence central to the interferometric measurements. However, after the spin-dependent momentum transfer, spin-motion separation develops necessarily in an interferometry sequence for sensing the spatial-dependent interactions, including those due to any spatially varying $\delta_{\rm com}$. A standard technique to counter the inhomogenuous $\delta_{\rm com}$ broadening relies on introducing additional sidebands to the $\mathcal{E}_{1,2}$ pulses with opposite Stark shifts~\cite{Asenbaum2020a}, or, by taking advantage of nanosecond operations, to fire additional phase-trimming pulses in the time domain~\cite{Qiu2022}. In the former case, the bias angle $\theta_b$ can be modified by the additional sidebands, which should be included during the CP-BR optimization.

Previously, high speed spinor matterwave control at the nanosecond level was mostly considered for manipulating microscopically confined ions~\cite{Garcia-Ripoll2003,Duan2004, Mizrahi2013, Heinrich2019}, typically involving THz-level single photon detuning $\Delta$ in the case of Raman excitation. By operating at tens of GHz detuning $\Delta$, our proposal has the obvious advantage of reducing the laser intensity $I\propto \Delta$ requirement for the Raman control, thereby supporting rapid control of macroscopic samples even with milli-Watt level laser power~\cite{Qiu2022}. More generally, the CP-BR method supports a suitable single-photon detuning $\Delta$ to balance the optical power efficiency with the requirements on the control speed and/or the suppression of excited-state dynamics. In addition to supporting the quantum-enhanced atom interferometry technology~\cite{Szigeti2020, Wu2020, Greve2021, Anders2021} for free-space samples, the CP-BR method can be particularly useful for tailoring LMT-enhanced control of macroscopic samples moving at high speeds, such as for interferometric rotation sensing~\cite{Gustavson2000} and nano-lithography~\cite{Johnson1998} with lightly collimated, high-flux atomic beams. To this end, we anticipate further developments of wideband pulse shaping techniques~\cite{He2020a,Wang2022d} for generating powerful, arbitrarily shapeable nanosecond pulses for spinor matterwave control of increasingly large atomic samples.

\section*{Acknowledgements}
We acknowledge support from National Key Research Program of China under Grant No. 2022YFA1404204, and from NSFC under Grant No.~12074083. H.Y. acknowledges partial support from the Research Grants Council of Hong Kong under Grant No. 14307420, 14308019,14309022 and the hospitality of Shenzhen International Quantum Academy.

\bibliography{sdkexpt, nfc}

\begin{thebibliography}{74}
\bibitem[{foo({\natexlab{b}})}]{foot:phase}
\bibinfo{note}{Practically, the uniform intensity requirement can be reasonably
  well satisfied for atomic samples with size $\sigma\ll w$ within the Rayleigh
  range of the $\mathcal{E}_{1,2}$ Gaussian profile with $w\gg\lambda$.}
\end{thebibliography}
\bibliographystyle{apsrev}

\pagebreak
\clearpage
\widetext
\begin{center}
\textbf{\large Supplemental Materials}
\end{center}
\setcounter{equation}{0}
\setcounter{figure}{0}
\setcounter{table}{0}
\setcounter{page}{1}
\setcounter{section}{0}
\makeatletter
\renewcommand{\theequation}{S\arabic{equation}}
\renewcommand{\thesection}{S\arabic{section}}
\renewcommand{\thefigure}{S\arabic{figure}}
\renewcommand{\bibnumfmt}[1]{[S#1]}

\setcounter{section}{0}

\twocolumngrid

\section{Full numerical model}\label{sec:num}

\begin{figure}[htbp]
    \centering
    \includegraphics[width=1\linewidth]{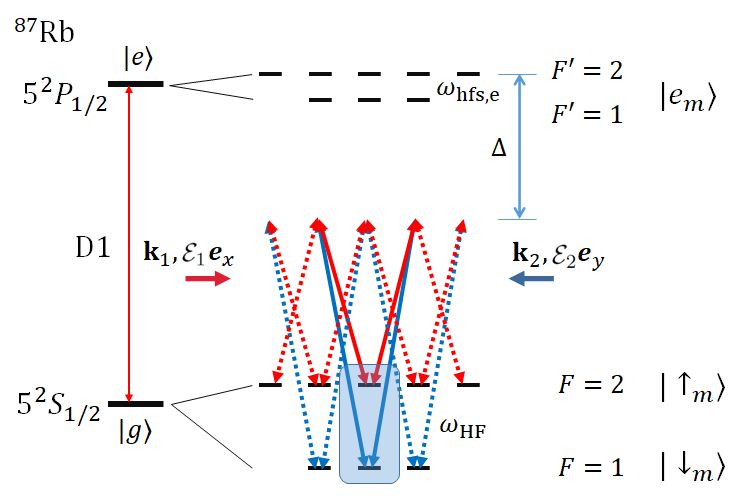}
    \caption{Schematic diagram for the full $^{87}$Rb D1 model. We consider counter-propagating ${\bf k}_{1,2}$ and linear${\perp}$linear polarization, with the quantization axis along ${\bf e}_z$~\cite{Qiu2022}. At $\Delta\gg \omega_{{\rm hfs},e}$, the hyperfine dynamics is decomposed into those within $\{\ket{\uparrow_m},\ket{\downarrow_m}\}$ sub-spins ($m=-1,0,1$), with suppressed $m$-changing leakages (Eq.~(\ref{eq:leakrabi})). The $\{\ket{\uparrow}, \ket{\downarrow}\}=\{\ket{\uparrow_0}, \ket{\downarrow_0}\}$ sub-spin for Fig.~\ref{fig:intro}(e,f), Fig.~\ref{fig:optim-pulsearea} is highlighted. }
    \label{fig:level}
\end{figure}

The numerical simulation in this work is based on the full light-atom interaction Hamiltonian on the D1 line of $^{87}$Rb, as schematically illustrated in Fig.~\ref{fig:level}, with the 5$P_{1/2}$ hyperfine splitting $ \omega_{{\rm hfs},e}=2\pi\times 814.5$~MHz~\cite{Steck2010} adjusted to zero when necessary.

In Fig.~\ref{fig:level}, the Zeeman-degenerate hyperfine states are labeled as $\ket{e_m},\ket{\uparrow_m},\ket{\downarrow_m}$ with magnetic quantum number $m=m_F$ respectively. We consider counter-propagating laser pulses, $\vo{E}_{1,2}={\vo{e}}_{1,2}\mathcal{E}_{1,2} \me^{i\qty(\vb{k}_{1,2}\cdot\vb{r}-\omega_{1,2}t)}+c.c.$ with perpendicular linear ${\bf e}_{1,2}$~\cite{Qiu2022}, shaped slowly-varying amplitudes $\mathcal{E}_{1,2}(\vb{r},t)$, and Raman-resonant carrier frequency difference $\omega_2-\omega_1=\omega_{\rm HF}=2\pi\times 6.835~$GHz~\cite{Steck2010} to address the ground-state atom. Following the discussion in the main text, the effective, non-Hermitian Hamiltonian is 
    \begin{equation}
        \begin{split}
            H_{\rm eff}(\vb{r},t) = &\hbar\sum_{e,l}\qty(\omega_{e}-\omega_{e0}-i\Gamma_e/2)\sigma^{e_le_l} + \\
            &\hbar\sum_{c=\uparrow,\downarrow,m}\qty(\omega_{c}-\omega_{g0})\sigma^{c_mc_m}+\\
            & \frac{\hbar}{2}\sum_{c={\uparrow,\downarrow}}\sum_{e,m,l}\Omega^{j}_{c_me_l}(\vb{r}, t)\sigma^{c_me_l} + {\rm h.c.}
        \end{split}\label{eq:HD1}
    \end{equation}
Here $\omega_{e0}, \omega_{g0}$ are decided by the energy of reference level in the excited and ground state manifolds respectively, chosen as the top hyperfine levels in this work. The laser Rabi frequencies,
    \begin{equation}
        \begin{split}
            \Omega^{j}_{e_l \uparrow_m\qty(\downarrow_n ) } (\vb{r},t)&\equiv -\frac{\mel{e_l}{\vb{d}\cdot\vo{e}_j\mathcal{E}_j(\vb{r},t)}{\uparrow_m\qty(\downarrow_n )}}{\hbar},
        \end{split}
    \end{equation}
  are accordingly written in the $\omega_{e0,g0}$ frame under the rotating wave approximation.     The $\vb{d}$ is the atomic electric dipole operator. The  $\sigma^{\uparrow_m e_l} = \op{\uparrow_m}{e_l}$, $\sigma^{e_l \uparrow_m} = \op{e_l}{\uparrow_m}$ are the raising and lowering operators between states  $|\uparrow_m\rangle$ and $|e_l\rangle$. Similar $\sigma$ operators are defined for all the other $|\uparrow_m\rangle$, $|\downarrow_n\rangle$ and $|e_l\rangle$ state combinations.   
  In according to the discussions in the main text, Eq.~(\ref{eq:HD1}) can also be expressed as $H_{\rm eff}=H_0-i\hat\Gamma/2$, with $\hat \Gamma=\Gamma \sum_{e,l}\sigma^{e_le_l}$, $\Gamma=1/(27.7~{\rm ns})$ to be the D1 linewidth~\cite{Steck2010}.  
 
Numerical evaluations of $\tilde U$ and the Eq.~(\ref{eq:gatef}) gate fidelity $\mathcal{F}$ in the main text follow the ref.~\cite{Qiu2022} recipe based on the Eq.~(\ref{eq:HD1}) Hamiltonian here. In particular, the radiation damping is reflected in the decreasing norm of the wavefunction $|\psi\rangle$~\cite{Dum1992}. During the gate fidelity evaluation, since any spontaneous emission is associated with complete decoherence, we only need to consider the non-hermitian evolution without any quantum jump~\cite{Dalibard1992,Carmichael1993}.

In the numerical simulations for Fig.~\ref{fig:optim-pulsearea}, Fig.~\ref{fig:optim2d} in the main text, we smooth the rising and falling edges of the area $\mathcal{A}$ square pulse. For the purpose, the pulse duration is first elongated from $\tau_c=m_{\mathcal{A}}\tau_{\pi}$ to $\tau_c=(\pi/2-1+m_{\mathcal{A}})\tau_{\pi}$. Next, the first and last $\tau_{\pi}/2$ are expanded into $\tau_{\pi}\pi /4$ durations, to form the sine-shaped rising and falling edges, thereby slowly ramp $|\Omega_{\rm R}|$ from zero to its maximum. Finally, the edge-smoothed pulse is divided into $N$ equal-area parts to associate with phase profile $\{\varphi_j\}$ of interest.

\section{Reduction to 2-level model}
We refer readers to ref.~\cite{Qiu2022} for the reduction from Eq.~(\ref{eq:HD1}) to the Eq.~(\ref{eq:H1}) spinor Hamiltonian in the main text for the spinor $\{\ket{\uparrow}=\ket{\uparrow_m}, \ket{\downarrow}=\ket{\downarrow_m}\}$ defined on a pair of hyperfine Zeeman sub-levels (Fig.~\ref{fig:level}). Here we generally note that with a slight rotation of the $H_0$ basis and by expressing (after a hyperfine rotating wave transformation) 
\begin{equation}
    H_{\rm eff}=H+V'-i\hat \Gamma/2,
  \end{equation}
then the $V'$ term includes all the unitary corrections from the full model, including the $m-$sensitive light shifts and couplings. For the linearly polarized ${\bf e}_{1,2}$ that only drives the $\Delta m=\pm 2$ leakages among Zeeman sublevels~\cite{Happer1972}, the leakages are associated with a coupling strength:
\begin{equation}\label{eq:leakrabi}
    \Omega^{\pm 2} = \mathcal{O}\qty(\frac{\omega_{\rm hfs,e}}{\Delta})\Omega_{\rm R}.
\end{equation}
Therefore, increasing $\Delta$  suppresses the spin leakage. Other terms in $V'$ corrections are at least $1/\Delta^2$-suppressed too, except for the ($m$-insensitive) $\delta_{\rm com}$ common shift as being discussed in Sec.~\ref{sec:conclusion}. In the large $\Delta$ limit so that the $V'$ contribution to atomic state dynamics vanishes, the decay loss by $\hat \Gamma$ is decided by the linewidth of the instantaneous ``dressed'' ground states, as discussed in the main text with Eq.~(\ref{eq:Isp}).



\section{Biased rotation}\label{sec:BRA}
Following Eq.~(\ref{eq:HD1}), we consider a spinor $|\{\ket{\up}\rangle=\ket{\uparrow_m}, \ket{\dn}=\ket{\downarrow_m}\}$ defined on a pair of ground state hyperfine Zeeman-sublevels. As depicted in Fig.~\ref{fig:intro}~(a), resonant Raman coupling is induced by $\Omega_{1,2}$ to coherently couple $\ket{\up}$ and $\ket{\dn}$, forming a effective spin-1/2 system subjected to the Eq.~(\ref{eq:H1}) Hamiltonian. The 2-photon shift by Eq.~(\ref{eq:shift}) is $\delta = \delta_{\up}-\delta_{\dn}$ with 
    \begin{equation}
        \begin{split}
            \delta_{\up} &= \frac{\abs{\Omega_{1}}^2}{4\Delta} + \frac{\abs{\Omega'_{2}}^2}{4\qty(\Delta+\omega_{\rm HF})}\\
            \delta_{\dn} &= \frac{\abs{\Omega_{2}}^2}{4\Delta} + \frac{\abs{\Omega'_{1}}^2}{4\qty(\Delta-\omega_{\rm HF})}
        \end{split}
    \end{equation}
Notice while $\Omega_{1,2},\Omega'_{1,2}\propto\sqrt{I_{1,2}}$, the relative strengths between $\Omega_{1,2}$ and $\Omega'_{1,2}$ are determined by the associated dipole transition matrix elements. Here, we consider the example of $I_1=I_2$ when driving the spinor defined on the D1 line so that  $|\Omega_{1,2}|=|\Omega'_{1,2}|=\Omega$. In this case,
    \begin{equation}
        \begin{split}
            \delta = \delta_{\up} - \delta_{\dn} &= \frac{\abs{\Omega}^2}{4\qty(\Delta+\omega_{\rm HF})} - \frac{\abs{\Omega}^2}{4\qty(\Delta-\omega_{\rm HF})}\\
            &= -\frac{\abs{\Omega}^2\omega_{\rm HF}}{2\qty(\Delta+\omega_{\rm HF})\qty(\Delta-\omega_{\rm HF})},
        \end{split}
    \end{equation}
with a bias ratio
    \begin{equation}
        {b} = \tan{\theta_{b}} = \frac{\delta}{\Omega_{R}} = -\frac{\Delta\omega_{\rm HF}}{\qty(\Delta-\omega_{\rm HF})\qty(\Delta+\omega_{\rm HF})}.
    \end{equation}

At an arbitrary pulse area $\mathcal{A}$, the SU(2) evolution is described by a propagator
    \begin{equation}\label{eq:propagtor-single}
        U(\mathcal{A},b, \varphi) = \mathbf{1}\cos\frac{\tilde{\phi}}{2} - i\sin\frac{\tilde{\phi}}{2}\frac{\sigma_x\cos\varphi+\sigma_y\sin\varphi + b\sigma_z}{\sqrt{1+b^2}}
    \end{equation}
    with rotation angle $\tilde{\phi} = \sqrt{1+b^2}\mathcal{A}$. As illustrated in Fig.~\ref{fig:intro} (b) in the main text, due to the finite $b$, the available rotating axes $n$ are biased from the equator by a fixed angle $\theta_{b}$, for which we call the associated SU(2) controls as ``biased rotation'' (BR).


\section{GRAPE optimization}\label{sec:numoptim}
    We use a gradient-based optimization algorithm called GRAPE (GRadient Ascent Pulse Engineering)~\cite{Khaneja2005} to optimize the target gates. 
    Specifically, for each phases $\varphi_i$ at each iteration, its gradient is
    \begin{equation}
        g_i = -\pdv{\mathcal{F}}{\tilde{U}}\pdv{\tilde{U}}{\varphi_i}.
    \end{equation}
The ${\partial\mathcal{F}}/{\partial\tilde{U}}$ is obtained from Eq.~\eqref{eq:gatef} directly, while the second term is
    \begin{equation}
        \pdv{\tilde{U}}{\varphi_i} = \tilde{U}_N\cdots \tilde{U}_{i+1}\pdv{\tilde{U}_i}{\varphi_i}\tilde{U}_{i-1}\cdots\tilde{U}_1.
    \end{equation}
 The $\tilde{U}_j$s are evaluated by Eq.~\eqref{eq:propagtor-single}. We further have \begin{equation}\label{eq:propagtor-grad-single}
        \pdv{\tilde{U}}{\varphi} = \mathbf{1}\cos\frac{\tilde{\phi}}{2} - i\sin\frac{\tilde{\phi}}{2}\frac{\sigma_y\cos\varphi-\sigma_x\sin\varphi}{\sqrt{1+b^2}}.
    \end{equation}

During the optimization for the $\mathcal{A}$-error resilience, the gradients are averaged over a list of errors $\qty{\varepsilon_\mathcal{A}}$
    \begin{equation}
        \overline{g_i}(\mathcal{A},b;\{ \varphi_j \})=\left\langle g(\mathcal{A}+\varepsilon_{\mathcal{A}}, b;\{ \varphi_j \})\right\rangle_{\{ \varepsilon_{\mathcal{A}}\}}.
    \end{equation}
The analytical evaluation of the propagators and gradients at SU(2) level helps to improve the computational efficiency and precision for the GRAPE optimization.



\section{$\pi/2$ gate}\label{sec:pi2}

\begin{figure}[htbp]
    \centering
    \includegraphics[width=1\linewidth]{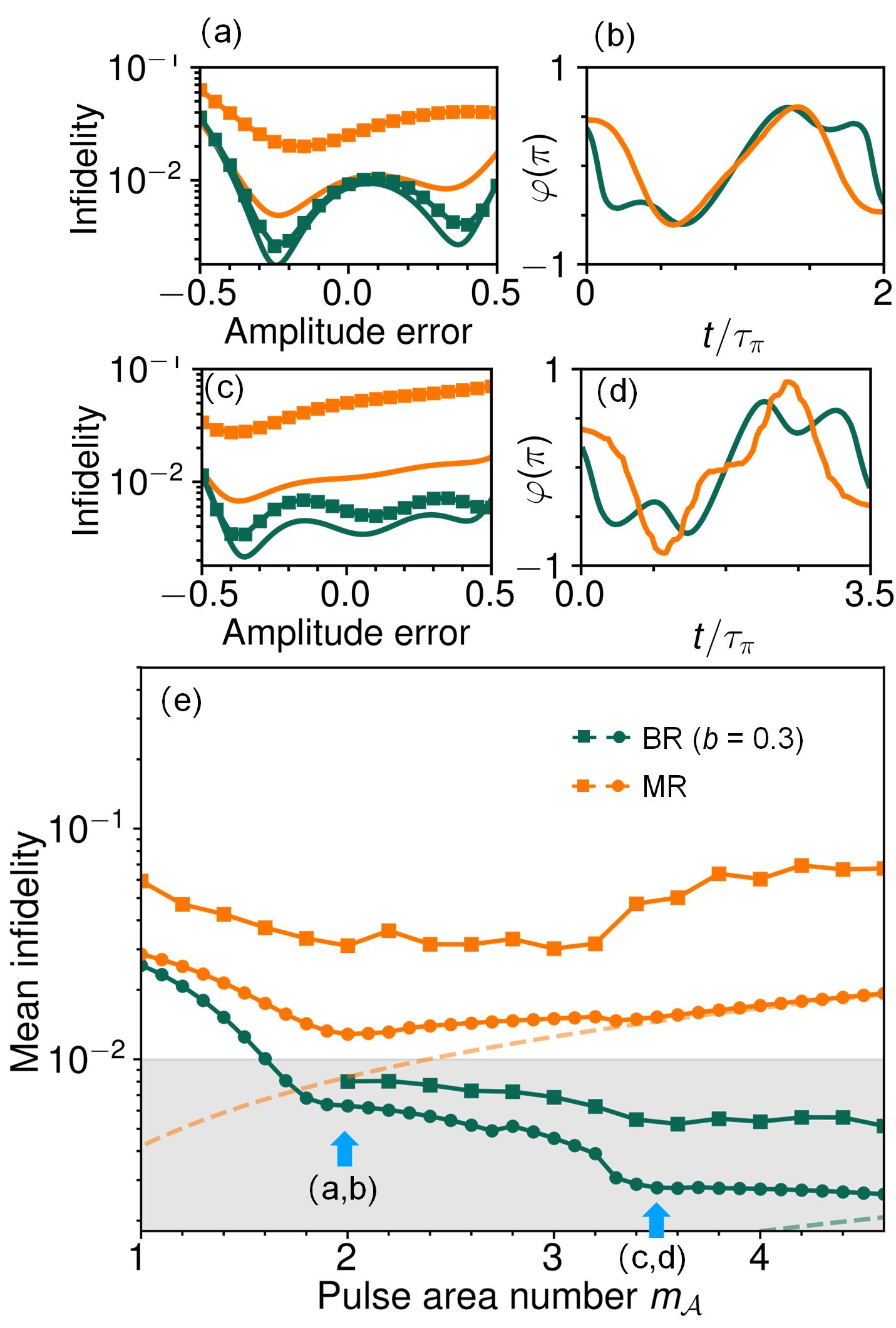}
    \caption{CP-BR for $R_x(\pi/2)$. The Fig.~(a,b) corresponds to Fig.~(a,i) in the main text, but with $\mathcal{A}=2\pi$. The Fig.~(c,d) corresponds to Fig.~(a,ii) in the main text, but with $\mathcal{A}=3.5\pi$. These pulse areas are marked with blue arrows in Fig.~(e) near the plateaus after the step-wise improvements to $\overline{\mathcal{I}}$ occur.} 
    \label{fig:pi2}
\end{figure}

Our discussions of CP-BR in the main text exploits the $R_x(\pi)$ example. In this section, we present an additional example of $R_x(\pi/2)$ with amplitude-error resilience. The results are presented in Fig.~\ref{fig:pi2}. Similar to the $R_x(\pi)$ case, the amplitude deviation uniformly samples $\varepsilon_{\mathcal{A}}\in(-0.5,0.5) \mathcal{A}$.  The bias ratio is again set as  $b=0.0$ (MR, with $\Delta=0.5\omega_{\rm HF}$) and $b=0.3$ (BR, with  $\Delta=3.6\omega_{\rm HF}$). The $\overline{\mathcal{I}}_{\rm opt}(\mathcal{A})$ vs $\mathcal{A}$ for the case of CP-BR (green line and symbols) show  step-wise features, similar to the $R_x(\pi)$ case in Fig.~\ref{fig:optim-pulsearea}. Comparing to the $R_x(\pi)$ results, here the ``plateau'' pulse area $\mathcal{A}$ following significant improvements to $R_x(\pi/2)$ ($m_\mathcal{A}=2$ and $m_\mathcal{A}=3.5$ in Fig.~\ref{fig:pi2}) are approximately $\pi/2$ less, which is somewhat expected as the desired $\pi/2$ rotating angle of $R_x(\pi/2)$ is $\pi/2$ less than that for $R_x(\pi)$. Similar to the $R_x(\pi)$ results in Fig.~\ref{fig:optim-pulsearea}a, the performance of the full-model $\overline{\mathcal{I}}$ degrades more substantially for the MR case, due to the more severe spin-leakage as being discussed in the main text. The apparently more severe BR-$\overline{\mathcal{I}}$ degradation is actually due to the improved fidelity for the $R_x(\pi/2)$ at the SU(2) level here (green disk symbols), compared to the $R_x(\pi)$ case, which makes the different more apparent. 

\sectionmain{Non-perfect implementation}\label{sec:robust}
    \begin{figure}[htbp]
        \centering
        \includegraphics[width=1\linewidth]{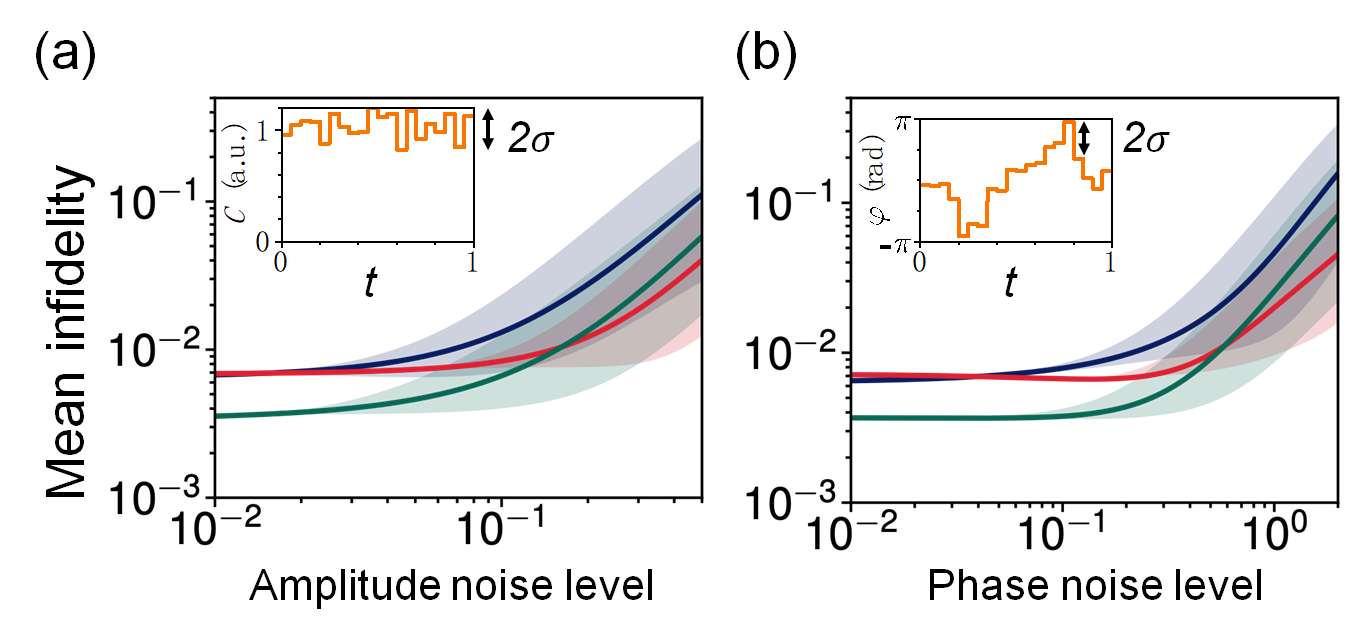}
        \caption{Performance of CP-BR for $R_x(\pi)$ with $b=0.3$ with non-perfect, random amplitude and phase noises distributed among the $N$ sub-pulses. The red, blue and green denote $(m_\mathcal{A}, N) = (4,20), (4,80), (5.5, 80)$ respectively. (a) $\overline{\mathcal{I}}$ as those in Fig.~\ref{fig:optim-pulsearea}a vs random amplitude noise. (b)  $\overline{\mathcal{I}}$ as those in Fig.~\ref{fig:optim-pulsearea}a vs random phase noise.}
        \label{fig:robust-precision} 
    \end{figure}

    In Fig.~\ref{fig:optim-pulsearea} in the main text and Fig.~\ref{fig:pi2} here we have shown that CP-BR achieve high-fidelity control ($\overline{\mathcal{F}}> 99\%$)  in presence of $\varepsilon_{\mathcal{A}}\in(-0.5,0.5) \mathcal{A}$ amplitude errors. 
    However, in real experiments, one can never implement CP with perfect amplitudes and phases as desired.
    To test the robustness of our CP solutions against non-perfect implementations, we consider the $R_x(\pi)$ example in Fig.~\ref{fig:optim-pulsearea} for a case study, by evaluating the CP-BR performance in presence of random noise among the sub-pulses. For each test, the random amplitude deviations uniformly sample between $\pm \sigma$, so that the relative amplitude of each sub-pulse is between $\tilde{C}_i = (1\pm\sigma)C_i$. 
The average $\overline{\mathcal{I}}$ from 100 random tests is plot against the maximum noise level $\sigma$ in Fig.~\ref{fig:robust-precision}~(a) with solid lines, while still covering the  $\varepsilon_{\mathcal{A}}\in(-0.5,0.5) \mathcal{A}$ mean amplitude error. The shadings suggest the $90\%$ variance from these 100 tests. The red, blue and green lines corresponds to pulse areas and pulse number as $(m_\mathcal{A}, N) = (4,20), (4,80), (5.5, 80)$ respectively.
    For $m_\mathcal{A}=4$ and $N=20$, the non-perfect CP implementation starts to impact the control fidelities when the noise level is larger than $5\%$. From Fig.~\ref{fig:robust-precision} we also see that a larger pulse number $N$ supports a stronger tolerance to waveform imperfections. For $m_\mathcal{A}=4$ and $N=80$ case, the maximally allowed deviations can be as large as $10\%$ while maintaining the target fidelity. On the other hand, for large pulse area $\mathcal{A}$ so that the actual pulse area per sub-pulse increases, the maximally allowed random noise level decreases. The case of random phase noise is similar to the case of amplitude, as in
    Fig.~\ref{fig:robust-precision}~(b).  Practically, we note accurate waveforms with a precision better than $95\%$ can be programmed directly with wideband optical waveform generation technique developed recently~\cite{He2020a,Wang2022d} after moderate waveform calibration.
    


\end{document}